# Confining Brownian motion of single nanoparticles in an ABELtrap

Maria Dienerowitz[a], Thomas Heitkamp[a], Thomas Gottschall[b], Jens Limpert[b,c], and Michael Börsch [a,c]

[a]Single-Molecule Microscopy Group, Jena University Hospital, Friedrich Schiller University, Jena, Germany
[b]Institute of Applied Physics, Friedrich Schiller University, Jena, Germany
[c]Abbe Center of Photonics (ACP), Jena, Germany

**ABSTRACT**

Trapping nanoscopic objects to observe their dynamic behaviour for extended periods of time is an ongoing quest. Particularly, sub-100nm transparent objects are hard to catch and most techniques rely on immobilisation or transient diffusion through a confocal laser focus. We present an **A**nti-**B**rownian **EL**ectrokinetic trap (pioneered by A. E. Cohen and W. E. Moerner [1–7]) to hold nanoparticles and individual $F_oF_1$-ATP synthase proteins in solution. We are interested in the conformational dynamics of this membrane-bound rotary motor protein that we monitor using single-molecule FRET. The ABELtrap is an active feedback system cancelling the nano-object's Brownian motion by applying an electric field. We show how the induced electrokinetic forces confine the motion of nanoparticles and proteoliposomes to the centre of the trap.

*Keywords*: ABELtrap, Brownian motion, Nanoparticles, $F_oF_1$-ATP synthase

## 1. INTRODUCTION

The ABELtrap is a great tool to monitor single molecules and nano-objects in real time in solution without the need for tethering or exposure to excessive laser radiation. It is capable of trapping individual fluorophore molecules [8–22] - some of the tiniest building blocks of single-molecule biophysics. This is extremely valuable when observing biomolecules at work - objects particularly inaccessible, even by optical tweezers, due their small size and transparency. The ABELtrap is an active feedback trap counteracting the Brownian motion of a nano-object in solution by applying an appropriate electric field across the trapping area. A laser scanning system detects the nano-object and encodes its position onto a single-photon detector. The FPGA-based feedback algorithm calculates the required feedback voltages based on the estimated object's position to direct it back to the centre of the trap.

The focus of our work is the mechanochemistry of the $F_oF_1$-ATP synthase enzyme. $F_oF_1$-ATP synthase is a bidirectional motor protein. Subunits of this enzyme comprising a central stalk rotate with respect to the peripheral subunits during catalysis. We developed a single-molecule Förster resonance energy transfer (smFRET) approach to monitor internal subunit rotation. We attach two different fluorophores to the enzyme: one to a rotating and another one to a non-rotating subunit. smFRET allows us to continuously measure the distance between these two markers. Our previous confocal smFRET studies revealed distinct step sizes of the two rotary motors of the $F_oF_1$-ATP synthase. The direction of rotation reverses when the enzyme switches from ATP synthesis to ATP hydrolysis. Elastic properties of the central rotor subunits compensate the mismatch of step sizes [23–37]. Yet, the drawbacks of these initial measurements are short observation times and large intensity fluctuations inherent to single molecules freely diffusing through a confocal detection volume. Our first experiments using an ABELtrap to hold single enzymes in place during the smFRET measurements extended the observation time to hundreds of milliseconds and up to seconds [38–43]. We expect a significant increase in precision of distance measurements within the enzyme in the experiments to follow.

E-mail: maria.dienerowitz@med.uni-jena.de, http://www.single-molecule-microscopy.uniklinikum-jena.de

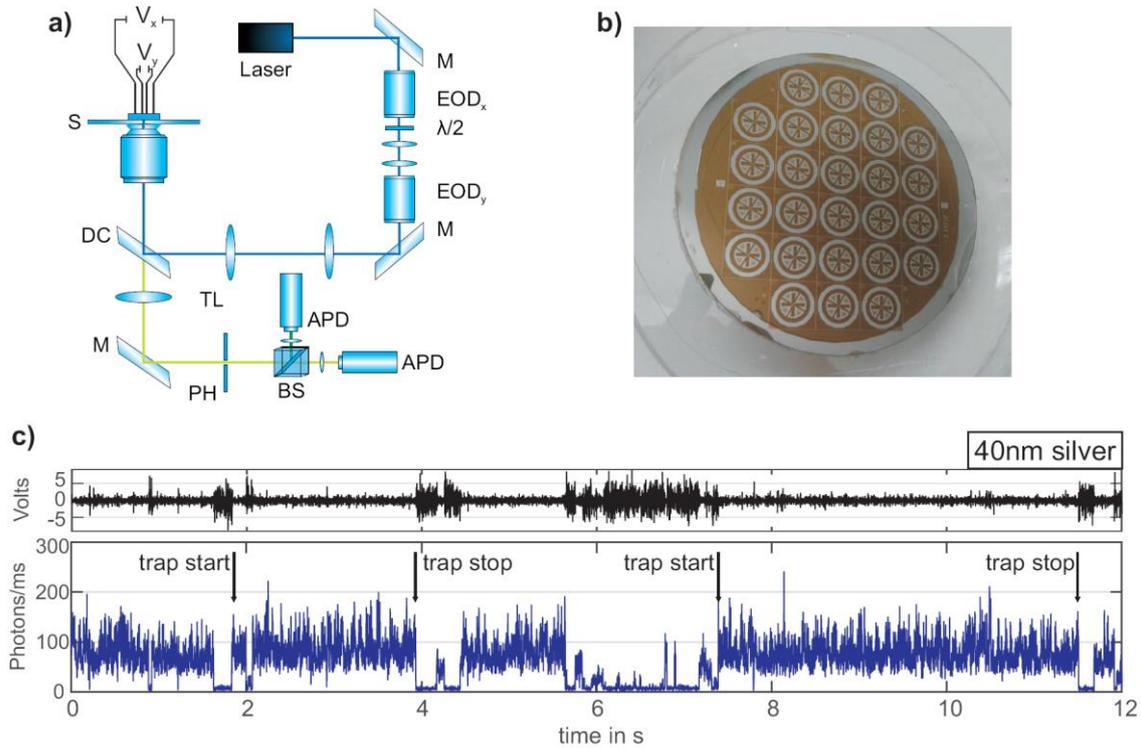

Figure 1. a) The excitation path of our ABELtrap setup consists of the excitation laser, mirrors (M), electro-optical deflectors (EOD), relay optics, a dichroic to couple the laser through the objective (O) into the sample (S). In the detection path, a tube lens (TL) focuses the detected photons through a 300µm pinhole (PH), dichroic beam splitter (BS) and bandpass filters onto single-photon counting avalanche diodes (APD). An FPGA tracks the position of the object of interest and applied feedback voltages (Vx, Vy) directly to the sample to counteract the object's Brownian motion. b) Photograph of the microfluidic chip mask wafer allowing us to produce 26 disposable chips in one process step. c) A typical photon intensity trace with the corresponding feedback voltages for trapping 40nm silver nanoparticles. An instant increase in photon counts as well as a decrease in applied voltages marks the start of a trapping event.

An inherent disadvantage of fluorophores is photobleaching: they eventually stop emitting photons. This marks the end of a trapping event since the ABELtrap does not have any other means to detect the object of interest. While the quest for better fluorophores is ongoing, engineered nanoparticles such as nanodiamonds [44–46] have proven promising. To that extend, we are interested in exploring very small metal nanoparticles in the ABELtrap exploiting the increased scattering cross section at the plasmon resonance.

## 2. MATERIALS AND METHODS

The experimental system is detailed in Fig. 1: two electro-optical beam deflectors (EOD, Conoptics) position the 491nm excitation laser (Calypso 50mW, Cobolt) at 40kHz in a predefined pattern to monitor the position of the nano-object within the trapping region. The 60x microscope objective (PlanApo N, oil, NA 1.42, UIS2, Olympus) focuses the laser into the microfluidic chip and images all emitted photons onto the single-photon counting avalanche diode (SPCM-AQRH 14, Excelitas). We use a nanopositioning stage (P-527.3CD, Physik Instrumente) for precise axial positioning of the sample chamber. The feedback algorithm controlling the electrokinetic trap runs on an FPGA (7852R PCIe, National Instruments) with an adapted version of the LABVIEW-based program by A. Fields & A. Cohen [12]. The microfluidic sample chamber consists of a PDMS chip (Sylgard 184 elastomer, Dow Corning) fabricated off a mask wafer [11, 41, 47] and bonded to a glass cover slip. The sample chamber consists of a 600nm high trapping region connected to four 80µm deep channels providing the sample solution and electric field. To facilitate future experiments with gold nanoparticles or alternative fluorophores we built an Ytterbium-based mode-locked fibre laser emitting at 515nm. The pulse duration of 40ps is adjusted with a narrowband fibre Bragg grating. To perform fluorescence lifetime measurements with this laser, we fine-tuned the repetition rate to 40MHz.

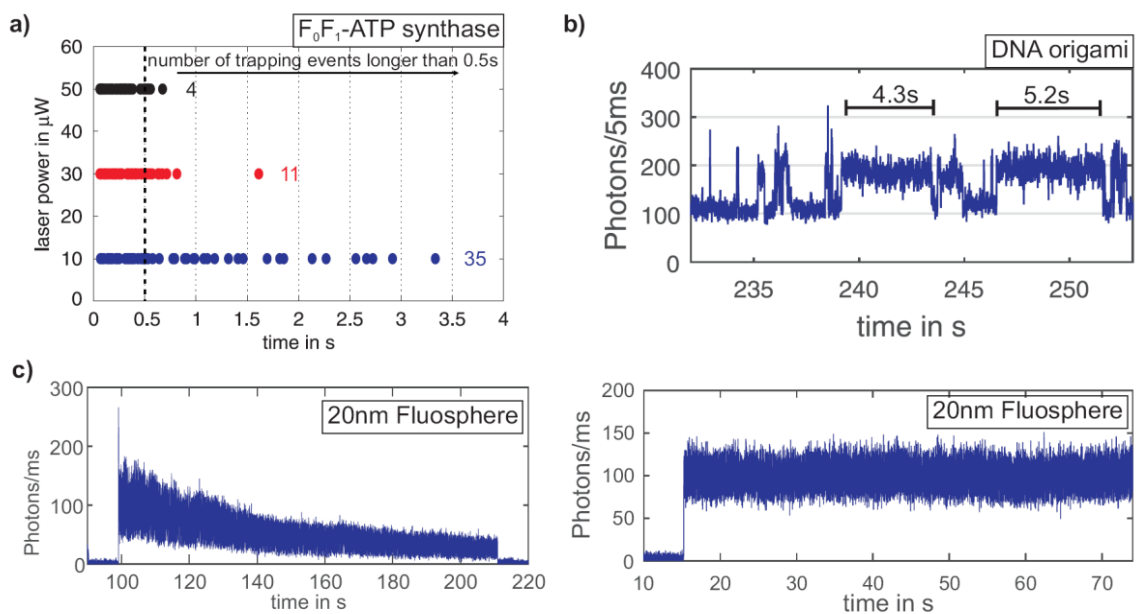

Figure 2. Reducing the excitation power in the ABELtrap is paramount to increase trapping and thus observation times. This is particularly true for single molecules as the bleaching of the attached fluorophore marks the end of a trapping event. a) The number of trapping events lasting longer than 0.5s when trapping single $F_oF_1$-ATP synthase in liposomes in the ABELtrap increases with decreasing excitation laser power from 50µW to 10µW. b) A typical photon count trace of trapped DNA origami labelled with one fluorophore shows trapping events lasting over 4s. c) 20nm Fluospheres are labelled with a multitude of fluorophores. Too high excitation powers (several µW) lead to photobleaching and thus a decrease in detectable photon counts (left) whereas the photon count rate remains constant (right) for low excitation powers (tens of nW).

To characterise and optimise the ABELtrap we investigate various kinds of nanoparticles and molecules. We purchased 40nm silver particles (BBInternational), 20nm 505/515 Fluospheres (Molecular Probes) as well as 100nm DNA origami squares labelled with a single Alexa Fluor 488 (GATTAquant). We purified the $F_oF_1$-ATP synthase from *Escherichia coli* and reconstituted the $F_oF_1$-*a*-mNeonGreen into preformed liposomes (120nm diameter) [48].

## 3. RESULTS

We are able to trap 20nm fluorescent particles for several minutes increasing the observation time by three orders of magnitude compared to free diffusion. Although Fluospheres are labelled with several fluorophores, minimising excitation power is still critical to avoid a decay of the photon detection rate due to photobleaching (see Fig. 2c). Photobleaching of individual fluorophores attached to single molecules ends the ABELtrap's ability to detect and trap the molecule of interest. Thus optimising the excitation power with respect to increasing the signal to background ratio is essential for extending the observation time with the ABELtrap. Fig. 2a) demonstrates the excitation power-dependence for trapping 120nm liposomes containing individual mNeonGreen-labelled $F_oF_1$-ATP synthase molecules. Further improvements in the detection path allow us to hold 100nm square DNA origami beyond 4s, lengthening the observation time for single molecules by a factor of 100 compared to free diffusion.

Detecting metal nanoparticles is inherently different since the emitted photons are not spectrally separated from the excitation laser. Acquiring a detectable signal requires us to take advantage of the increased scattering cross section close to the nanoparticle's plasmon resonance. Increasing the excitation power results in an equivalent increase in background noise, even more than in the fluorescent case, since we are not filtering out the excitation laser. Alternative approaches include spatial filtering or using the nanoparticle's plasmon luminescence [49–53].

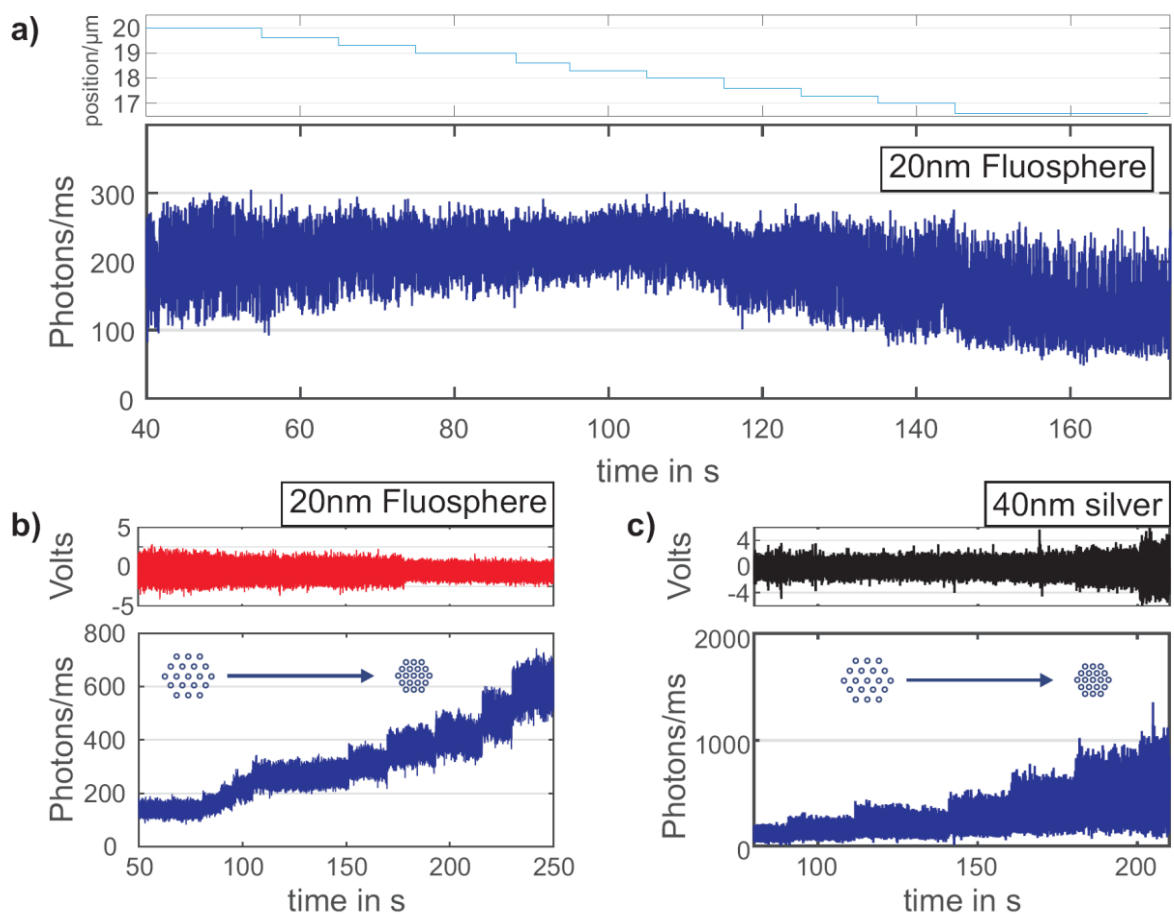

Figure 3. a) Precise axial positioning increases signal over background photon counts and reduces fluctuation of the photon count rate (90-110s). We started with the laser focus positioned below the trapping region and axially moving the stage down in 200-300nm steps. b) Reducing the size of the excitation laser pattern and thus the trapping region (d=2.8µm-1.4µm in 200nm steps) increases the signal over background ratio and confines the trapped 20nm Fluosphere to a smaller region. c) However this does not hold true for trapping 40nm silver nanoparticles since the detection relies on scattering rather than fluorescence and subsequently suffers from increased background noise.

We investigated the effects of axial positioning and varying trapping area size by adjusting the diameter of the laser scan pattern (see Fig. 3). Due to the shallow trapping region (depth 600nm), axial displacements of 200nm change the signal quality. Decreasing the trap size results in a stronger confinement and signal to background increase for fluorescently labelled spheres. However, for metal nanoparticles the fluctuation of the photon count rate increases with decreasing trap size without a reduction in the required feedback voltage that would indicate a tighter confinement. We attribute this to the previously discussed scattering-based detection scheme that ultimately raises the background noise without considerable signal gain when concentrating the excitation laser power to a smaller trapping region (see Figure 3c). We are capable of trapping 40nm metal nanoparticles for several minutes with as little as 5µW excitation laser power. Since the actual force holding the nanoparticles in place stems from the applied electric field and not the excitation laser, the ABELtrap avoids laser induced heating issues as observed in optical tweezers [54, 55].

## 4. OUTLOOK

The ABELtrap offers exciting prospects to hold and manipulate single molecules as well as nanoparticles even beyond fluorescently labelled objects. Furthermore it provides the means to extract the diffusion coefficient as well as the electrokinetic mobility from each individually trapped object. These nano-objects include chemically fuelled nanoswimmers, for example asymmetric platinum nanoparticles catalysing $H_2O_2$ decomposition. The ABELtrap is an ideal tool to study how individual nanoswimmers stochastically change their speed and direction of motion. It allows us to access the heterogeneities of individual catalytic activities of these inorganic nanomotors and their distributions. We also aim to further decrease the size of trappable metal nanoparticles.

Future biophysics work includes implementing smFRET in the ABELtrap to quantify subunit rotation and regulatory conformational changes within the $F_oF_1$-ATP synthase. Characterising other protein nanomachines such as ATP-driven transporters, ion channels or membrane receptors with smFRET requires detailed knowledge of the photophysics of the attached fluorophores. The ABELtrap allows us to optimise fluorophore observation times and improve the precision of smFRET distance measurements. Alternatively, replacing the currently used photobleachable fluorophores by inorganic nanoparticles has the potential to significantly extend the observation times of single biological nanomotors at work. Nanoparticles can be modified for specific attachment to the biological target molecule, for example using the biotin/streptavidin binding approach. As mentioned earlier, investigating gold nanoparticles requires adjusting the excitation laser wavelength to better match the plasmon resonance of gold.

**ACKNOWLEDGMENTS**

We are grateful for the continuous support by Alex Fields, Adam E. Cohen and W. E. Moerner. Maria Dienerowitz gratefully acknowledges funding from SPIE. We want to emphasize that Thorsten Rendler, Marc Renz and Anastasiya Golovina-Leiker (Stuttgart, Germany) have built previous EMCCD-based versions of an ABELtrap and tested the design of our PDMS microfluidics. Nawid Zarrabi and Monika Düser have built the AOD-based confocal ABELtrap (Stuttgart, Germany). This work was supported in part by the Thuringian State Government within its ProExcellence initiative (ACP2020) (to J. Limpert and M. Börsch) and DFG grant BO1891/16-1 (to M. Börsch).